\documentclass[a4paper]{jpconf}
\usepackage{graphicx}
\begin{document}
\title{Hydrodynamical Collapse of Neutron Stars due to Hadron-Quark Phase Transition}

\author{Guilherme F. Marranghello$^1$, Jos\'e Carlos N. de Araujo$^2$ and Oswaldo D. Miranda$^2$}

\address{$^1$Universidade Federal do Pampa, Bag\'e/RS, Brazil}
\address{$^2$Instituto Nacional de Pesquisas Espaciais, S\~ao Jos\'e dos Campos/SP, Brazil}

\ead{gfrederico.unipampa@ufpel.edu.br}

\begin{abstract}
We present studies of the collapse of neutron stars that undergo a hadron-quark phase transition. A spherical Lagrangian hydrodynamic code has been written. As initial condition we take different neutron star configurations taking into account its density, energy density and pressure distribution. The phase transition is imposed at different evolution times. We have found that a significant amount of matter on the surface can be ejected while the remaining star rings in the fundamental and first pressure modes.
\end{abstract}

\section{Introduction}

Black holes ($BHs$) and neutron stars ($NSs$) are certainly two major potential sources
of gravitational waves ($GWs$). Unlike $BHs$, whose gravitational waveforms are specified
essentially by their masses and angular momenta, the characteristics of the
gravitational emission from $NSs$ depend on the properties of the nuclear matter.

When nuclear matter is compressed to a sufficiently high density, it turns
into uniform three-flavor ($u$, $d$, and $s$)
strange quark matter (SQM), since it is expected that SQM may be more
stable than nuclear matter.
The deconfined quark matter
appears when the density is so high that the nucleons are
``touching'' each other.
At this point, when the number density of nucleons
$n\sim O(1\ {\rm fm}^{-3})$, the quarks lose their correlation
with individual nucleons and appear in a deconfined phase.
Since the density required for this to happen is not much higher than nuclear matter
density ($0.16\ {\rm fm}^{-3}$), the dense cores of neutron stars are
the most likely places where the phase transition to quark matter may
occur astrophysically.
It should be noted that strange quarks (in a confined phase)
could already exist in neutron stars with a hyperon core\cite{marranghello}.

In principle, the existence of a thin crust of normal
matter is possible at the surface of a strange star.
On the other hand, if SQM is metastable at zero pressure,
so that it is relatively more stable than nuclear matter only because
of the high pressure in the cores of neutron stars, then the final
products of the phase transition would be hybrid stars, which consist
of quark matter cores surrounded by normal matter outer parts.

The above arguments were made in \cite{lin} and those authors used polytropic equation of state to describe the complex nuclear matter structure. We proceed similar calculations with a realistic equation of state based on a field theoretical model. We also make use of a spherical Lagrangian hydrodynamical code instead of the Eulerian model used in \cite{lin}.

\section{Hydrodynamical Collapse}

The spherical Lagrangian hydrodynamic code
divides the star into several concentric shells. This number of shells
is the optimal combination for the variable time step and the number
of shells that gives the shortest processing time. Our results show
that 1600 shells are enough to model satisfactorily all the stars here
studied. To verify if the number of shells is an optimal choice we ran
a couple of models with 6400 shells, and we see that the results are the same
as for the models with 1600 shells. The shocks that appear are treated with the inclusion of the artificial viscosity of Richmeyer and Morton \cite{rich}.

The hydrodynamic equations that describe the dynamics of neutron stars are:

\begin{equation}
{\partial \rho\over\partial t}+{1\over r^2}{\partial\over\partial r}
(r^2\rho v)=0,
\end{equation}
the mass conservation equation written in spherical coordinates;
where $\rho$ is the density, $r$ the radial coordinate and
$v$ the velocity. In our code the continuity equation is calculated
directly from the grid. The specific volumes of the Lagrangian shells
give the density of the star.

The equation of motion is given by:

\begin{equation}
{D\vec v\over Dt}+{1\over\rho}\nabla P+\nabla\phi=0
\end{equation}
where the velocity is $\vec v$; the pressure of the nuclear matter is given by the model described in the next section.
The gravitational potential is $\phi$.
The energy equation is written as:

\begin{equation}
{DE\over Dt}={P\over\rho^2}{D\rho\over Dt}
\end{equation}
where $E$ is the thermal energy per gram.

\section{Neutron Star vs. Equation of State}

In our investigation we consider a
phenomenological QHD Lagrangian density\cite{taurines,ij1,ij2} which contains  the
fundamental baryon octet, the scalar meson fields $\sigma, f_0\/$
and the vector meson fields $\omega, \rho, \phi\/$. Additionally, it
contains the lightest charged leptons ($e$ and $\mu$) to allow for
charge neutrality,
\begin{eqnarray}
&{\cal L}&=\sum_{B = N,Y}[ \bar\psi_B i \gamma_\mu \partial^\mu \psi_B -
\bar\psi_B ( M_B - g_{\sigma_B}^{\star}\,\sigma
- g_{\sigma\ast B}^{\star} \,\sigma^\ast)\psi_B
- g_{\omega_B}^{\star} \bar\psi_B \gamma_\mu  \psi_B \omega^{\mu}
\nonumber \\ &-& g_{\phi_B}^{\star} \bar\psi_B \gamma_\mu \psi_B \phi^\mu
-\frac{1}{2} g_{\varrho_B}^{\star}
            \bar{\psi}_B \gamma_\mu\mbox{\bf$\tau$} \cdot
            \mbox{\bf$\varrho$}^\mu \psi_B]
-\frac{1}{4} \phi_{\mu\nu}\phi^{\mu\nu}
+ \frac{1}{2} m_\phi^2 \phi_\mu \phi^\mu
- \frac{1}{4}\omega_{\mu \nu}\omega^{\mu \nu}
\nonumber \\ &+&  \frac{1}{2} {m_\omega}^2 \omega_\mu \omega^\mu
           - \frac{1}{4}
              \mbox{\bf$\varrho$}_{\mu \nu} \cdot
              \mbox{\bf$\varrho$}^{\mu\nu}
          +\frac{1}{2} m_\varrho^2 \mbox{\bf$\varrho$}_\mu \cdot
              \mbox{\bf$\varrho$}^\mu
+\frac{1}{2}(\partial_\mu \sigma \partial^\mu \sigma-
             {m_\sigma}^2 \sigma^2)
\nonumber \\ &+&\frac{1}{2} (\partial_\mu\sigma^{\ast} \partial^\mu\sigma^{\ast}
-m_\sigma^{\ast 2} \sigma^{\ast 2})
+\sum_l \bar\psi_l \,[\,i\,\gamma_{\mu}\partial^{\mu} - m_l \,]\, \psi_l
\label{hyp10a}
\end{eqnarray}
where the parameterized coupling constants \cite{taurines} of the theory are defined as
\begin{eqnarray}
&g_{\sigma_B}^{\star} \equiv m_{B \alpha}^{\ast} \, g_{\sigma_B}\,, \,\,
g_{\sigma\ast_B}^{\star} \equiv m_{B \alpha}^{\ast} \,g_{\sigma^\ast_B}& \nonumber \\
&g_{\omega_B}^{\star} \equiv m_{B \beta}^{\ast} \, g_{\omega_B}&  \nonumber \\
&g_{\phi_B}^{\star} \equiv  m_{B\beta}^{\ast} \, g_{\phi_B}\,, \,
g_{\varrho_B}^{\star} \equiv m_{B\gamma}^{\ast} \, g_{\varrho_B}&
\label{hyp10b}
\end{eqnarray}
with
\begin{equation}
m_{B n}^\ast = \left( 1 + \frac{g_{\sigma _B} \sigma+g_{\sigma^{\ast}_B} \sigma^{\ast}
}{n\,M_B}\right)^{-n}\,, \,\,\, n = \alpha, \, \beta, \, \gamma\,.
\label{hyp11}
\end{equation}

For the quark matter the MIT Lagrangian \cite{grand75} was adopted.
Here we have assumed that the phase transition is of first order.
This last possibility is not favored energetically when the interface tension
between quarks and hadrons is important \cite{heiselberg93,heiselberg99}. Thus, the physical
conditions at deconfinement were estimated from Gibbs criteria.

\section{Numerical Results and Conclusions}

\indent
We have obtained some important results which seems to be in complete agreement with
those obtained by \cite{lin}. The equation of state is shown in Fig.1 for three different parameter $\alpha$ which leads to different hadron-quark phase transition pressure and density. Three stages of evolution are shown in Fig.2. The star radius shrinks and the central density is increasing during the collapse. In Fig.3 one can find the density evolution of each shell. The oscillation of the inner shells is clear in the final stage of evolution as it is the ejection of the external shells. The fourier transform is shown in Fig.4. This result shows a peak very close to the one shown
by \cite{lin}, at 3.1kHz, which shall correspond to the fundamental mode as well as other
peaks, corresponding to the first pressure modes. It is also easy to find that most of the
energy is released in the fundamental mode, as always claimed. In Fig.5 the radius and density evolution are plotted against time in this same figure.

Finally, after a future detection, one can try to fit the nuclear matter parameters to obtain the frequency of the detection, remembering that, if the frequency falls on the 3.0-3.4kHz bandwidth, the Brazilian Schenberg antenna will be able to identify the
direction of the source, as shown in \cite{mara}.

\ack{We would like to thank CNPq and Fapesp for financial support.}

\vskip 15pt

\begin{figure}[h]\hspace{2cm}
\includegraphics[width=30pc]{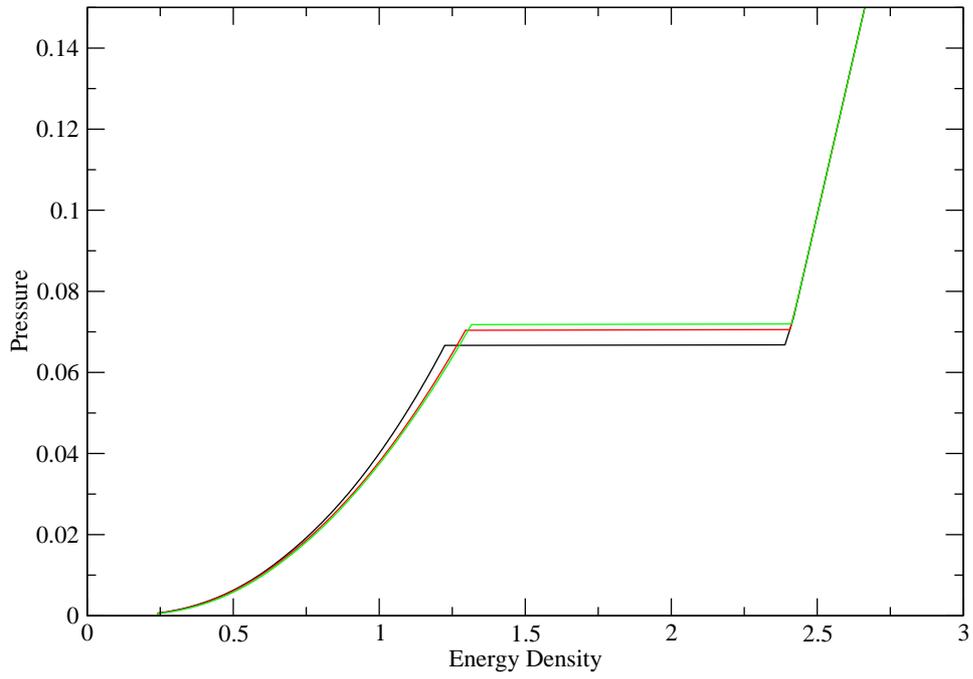}\vspace{0cm}
\caption{\label{label}Equation of state with hadron-quark phase transition for three different $\alpha$-parameter.}
\end{figure}

\begin{figure}[h]\hspace{2cm}
\includegraphics[angle=-90,width=30pc]{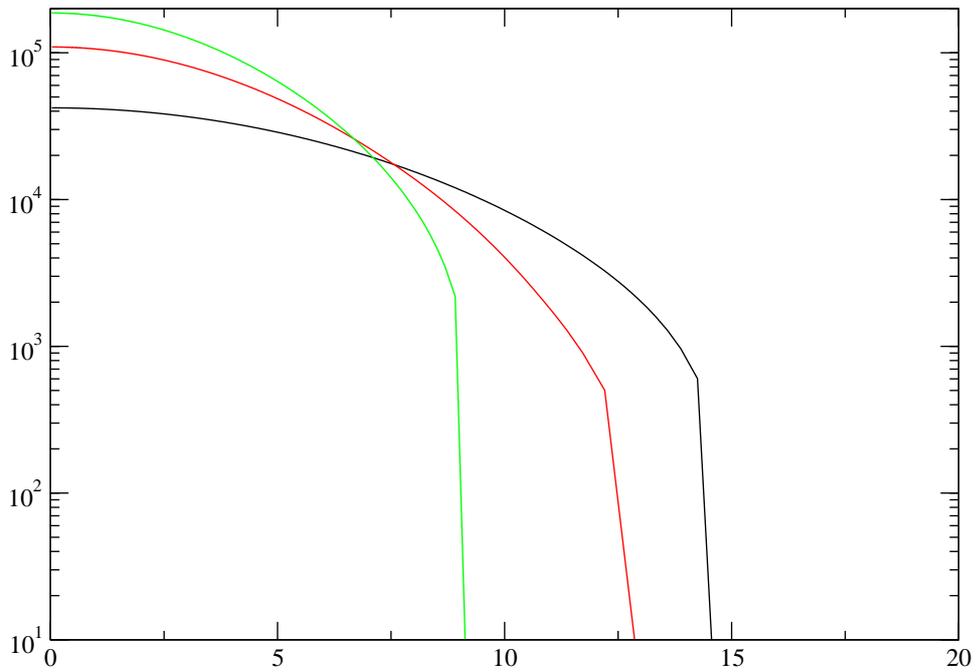}\vspace{1cm}
\caption{\label{label}The collapsed star structure at three different instants during the compression.}
\end{figure}

\begin{figure}[h]\hspace{2cm}
\includegraphics[angle=-90,width=30pc]{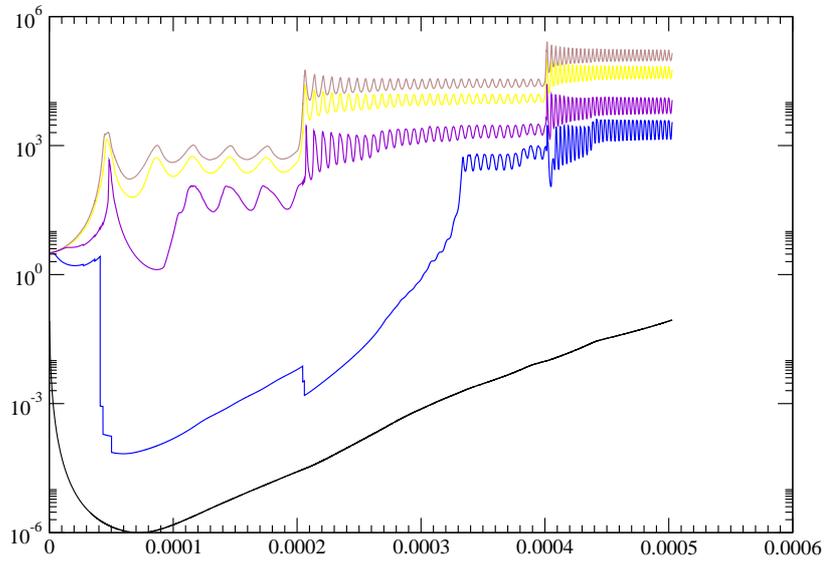}\vspace{0cm}
\caption{\label{label}Density shells evolution from collapse to oscillation modes. Time is expressed in seconds.}
\end{figure}

\begin{figure}[h]\hspace{2cm}
\includegraphics[angle=-90,width=30pc]{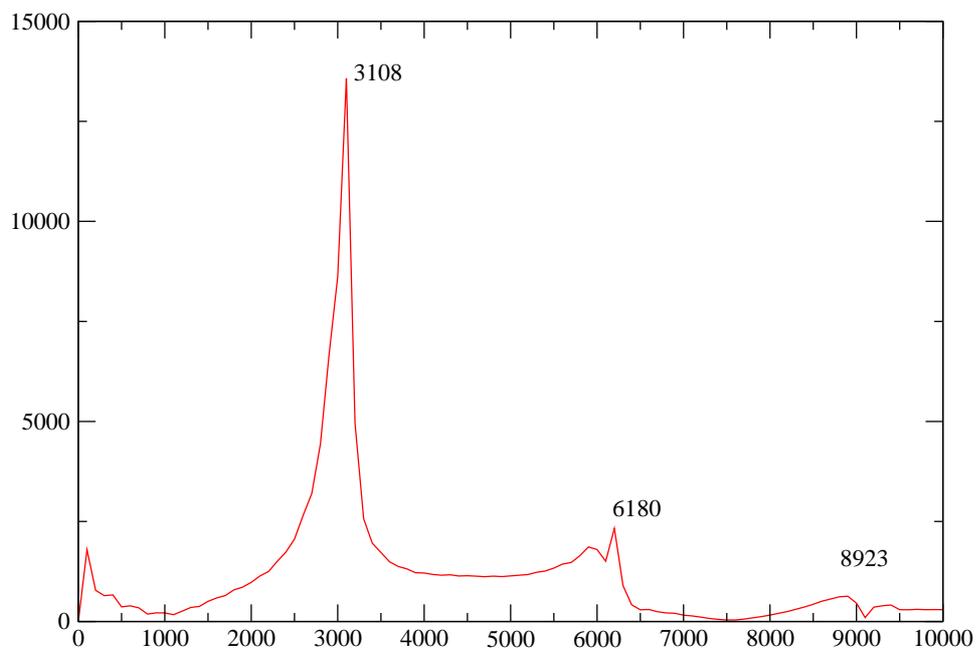}\vspace{1cm}
\caption{\label{label}Fourier transform of the oscillations. Frequency is written in kHz.}
\end{figure}

\begin{figure}[h]\hspace{1cm}
\includegraphics[width=35pc]{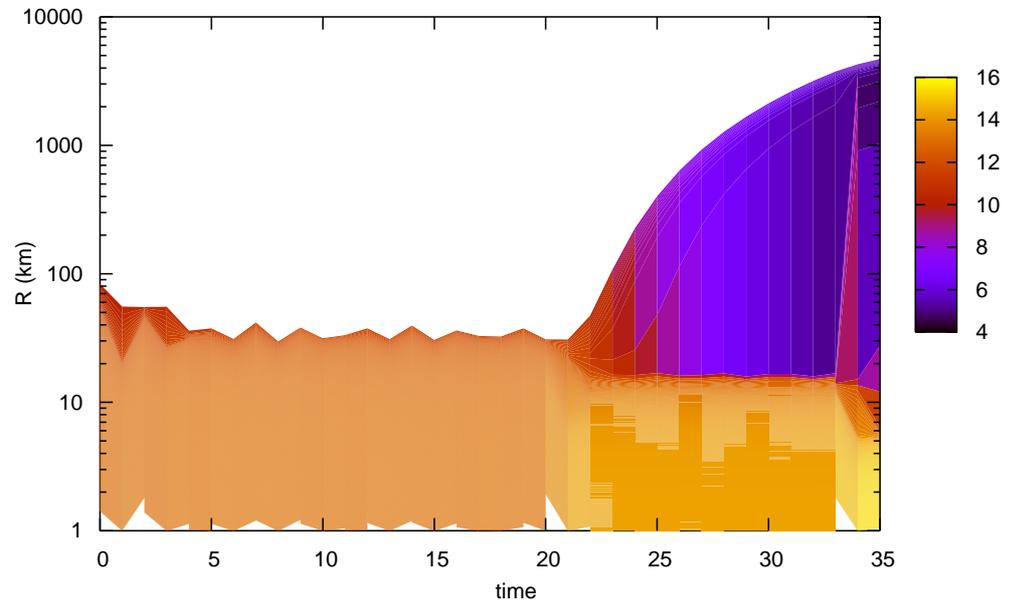}\vspace{0cm}
\caption{\label{label}Density evolution with mass ejection. Time is plotted in ms and the colors are the logarithm of energy density in $g/cm^3$.}
\end{figure}

\end{document}